\documentclass[%
 reprint,
superscriptaddress,
 amsmath,amssymb,
 aps,prl
]{revtex4-1}

\usepackage{graphicx}
\usepackage{dcolumn}
\usepackage{bm}
\usepackage{times}
\usepackage{tabularx}
\usepackage[usenames,dvipsnames]{xcolor}

\begin{document}

\title{Negative thermal expansion in transition-metal dicyanides: the hidden role of the underlying diamondoid framework} 

\author{Quentin Guéroult}
\affiliation{Inorganic Chemistry Laboratory, University of Oxford, South Parks Road, Oxford OX1 3QR, U.K.}%
\author{Johnathan M. Bulled}
\affiliation{Inorganic Chemistry Laboratory, University of Oxford, South Parks Road, Oxford OX1 3QR, U.K.}%
\author{Henry Patteson}
\affiliation{Inorganic Chemistry Laboratory, University of Oxford, South Parks Road, Oxford OX1 3QR, U.K.}%
\author{Chloe S. Coates}
\affiliation{Inorganic Chemistry Laboratory, University of Oxford, South Parks Road, Oxford OX1 3QR, U.K.}%
\author{Ronald I. Smith}
\affiliation{ISIS Facility, Rutherford Appleton Laboratory, Harwell Campus, Didcot OX11 0QX, U.K.}%
\author{Helen Y. Playford}
\affiliation{ISIS Facility, Rutherford Appleton Laboratory, Harwell Campus, Didcot OX11 0QX, U.K.}%
\author{David A. Keen}
\affiliation{ISIS Facility, Rutherford Appleton Laboratory, Harwell Campus, Didcot OX11 0QX, U.K.}%
\author{Andrew L. Goodwin}
\affiliation{Inorganic Chemistry Laboratory, University of Oxford, South Parks Road, Oxford OX1 3QR, U.K.}%

\date{\today}

\begin{abstract}
The transition-metal dicyanides M(CN)$_2$ (M = Zn, Cd) are amongst the most important negative thermal expansion (NTE) materials known, favoured for the magnitude, isotropy, and thermal persistence of the NTE behaviour they show. The conventional picture of the NTE mechanism in this family is one of correlated rotations and translations of M(C/N)$_4$ polyhedra acting to draw the diamondoid network of M--CN--M linkages in on itself. An implication of this mechanism is increased transverse vibrational motion of C and N atoms relative to the isotropic displacements of M atoms, which act as anchors. Here, we use a combination of neutron total scattering measurements and \emph{ab initio} calculations to reassess the vibrational behaviour of the M(CN)$_2$ family. We find that M, C, and N atoms all exhibit similar degrees of local thermal motion, such that the cyanide linkages behave as pseudo-springs connecting M$\ldots$M pairs. This interpretation leads us to uncover a `hidden' dispersion in the M(CN)$_2$ phonon dispersions, closely related to that of diamond and silicon themselves. By virtue of this mapping, a simple geometric model based on the competing energy scales of network stretching and flexing---long applied to interpret NTE modes in C and Si---turns out to capture the key NTE physics of M(CN)$_2$, especially at low temperatures. Our study highlights the potential insight gained by coarse-graining the complex lattice dynamics of framework materials in terms of what we call `framework modes'---the correlated distortions of the underlying network structure itself.
\end{abstract}

\maketitle

\emph{Introduction}---Most materials exhibit positive thermal expansion (PTE), with around a  0.3\% increase in volume for each 100\,K temperature rise \cite{Hidnert_1943,Krishnan_1979}. Certain materials exhibit negative thermal expansion (NTE) instead \cite{Evans_1999,Barrera_2005,Coates_2019}. These can be combined with conventional PTE materials to design composites with tuneable coefficients of thermal expansion (CTEs)---\emph{e.g.}\ for applications that require the dimensions of a material to be maintained over large temperature gradients \cite{Dow_1969,Sideridou_2004,Legero_2010,Takenaka_2012,Romao_2013,Chen_2015}. Isotropic NTE is also of particular fundamental interest because it can signal a variety of anomalous mechanical responses, including auxeticity \cite{Coudert_2015} and pressure-induced softening \cite{Pantea_2006,Fang_2013}.

\begin{figure}
    \centering
    \includegraphics{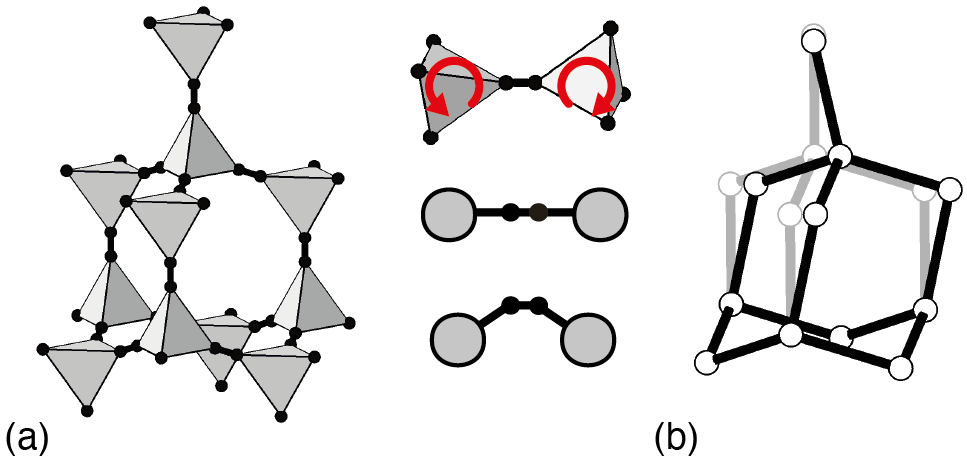}
    \caption{Two possible mechanisms that drive NTE in diamondoid frameworks. (a) Rigid-unit mode rotations of neighbouring coordination polyhedra---as proposed to occur in M(CN)$_2$ frameworks---give rise to local transverse displacements away from the M$\ldots$M axis and so shorten the M$\ldots$M distance if bond-lengths are preserved. (b) Acoustic (shear) modes in C and Si distort network angles but not bond-lengths, and couple to volume reduction accordingly.}
    \label{fig1}
\end{figure}

Zinc and cadmium cyanides, M(CN)$_2$ (M = Zn, Cd), exhibit particularly strong isotropic NTE \cite{OKeefe_1997,Kepert_2005,Chapman_2005, Zwanziger_2007}. Their negative volumetric expansion coefficients are more than double that of other well-known NTE materials such as ZrW$_2$O$_8$ \cite{Mary_1996}, and persist over large temperature ranges. The NTE mechanism of the M(CN)$_2$ family is usually described in terms of cooperative rotations of M(C/N)$_4$ tetrahedra [Fig.~\ref{fig1}(a)]. As rotation occurs, the C and N atoms of each M--CN--M linkage displace away from the local M$\ldots$M axis, giving rise to a `skipping-rope'-type motion that draws the structure in on itself \cite{Chapman_2005,Hibble_2013}. The many different ways in which polyhedral rotations can be combined within the diamondoid M(CN)$_2$ structure result in a set of essentially dispersionless NTE bands in the low-energy phonon spectrum. In addition to these rotational modes, two low-energy translational branches are also implicated in NTE, but the microscopic origin of these modes is less well understood \cite{Misquitta_2013}.

\begin{figure*}
    \centering
    \includegraphics{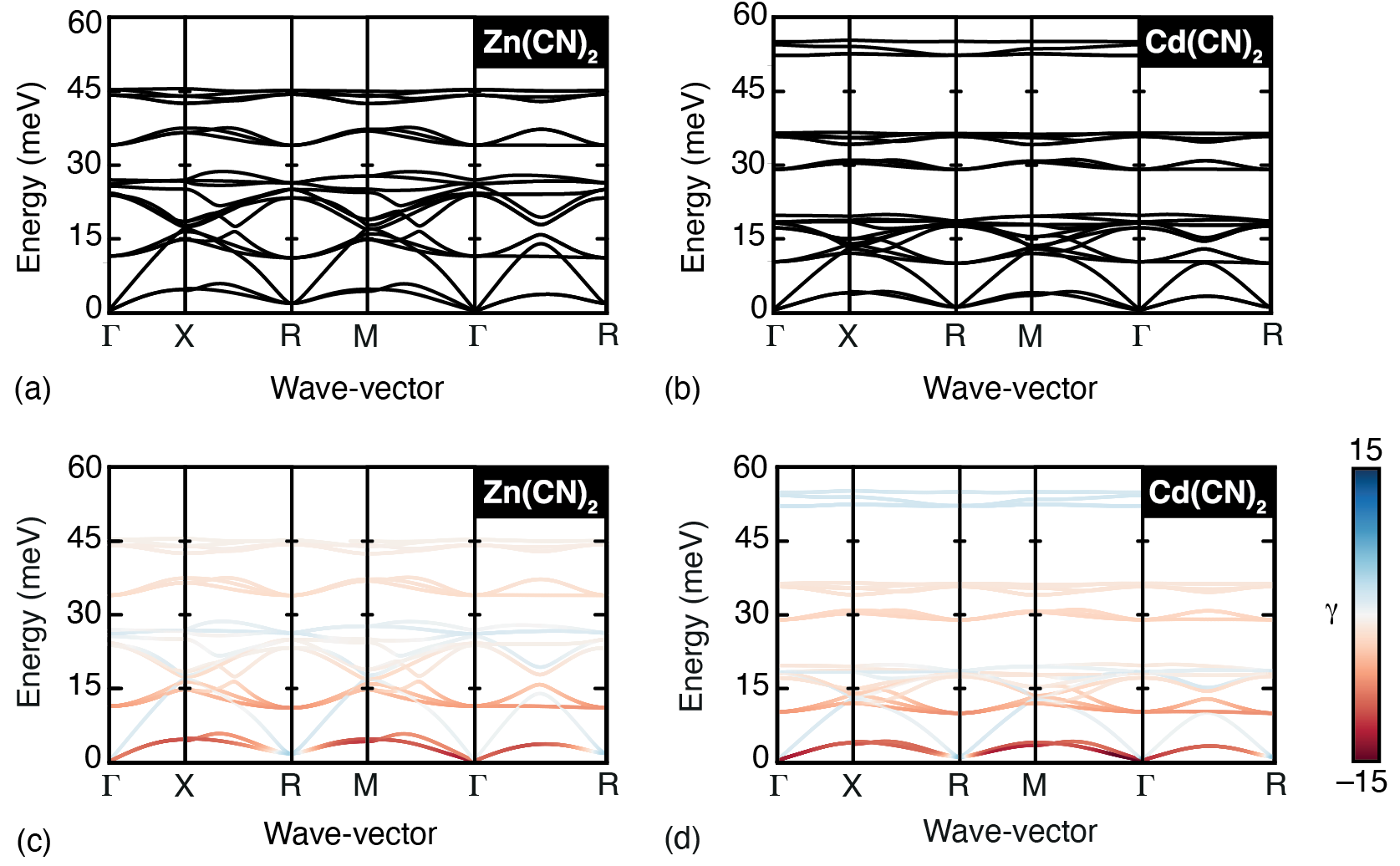}
    \caption{Four sets of M(CN)$_2$ dispersion curves, calculated using DFT. For (a) and (b), these are the dispersion curves for Zn(CN)$_2$ and Cd(CN)$_2$ in conventional representation; note that the cyanide stretching bands occur at very high energy and are not shown here. For (c) and (d), the same dispersion curves are coloured according to the corresponding value of the mode Gr\"uneisen parameter. The colour map used ranges from red ($\gamma=-15$) to blue ($\gamma = +15$), with white indicating a value of zero. The transverse acoustic modes are the branches with the most negative $\gamma$ (\emph{i.e.}\ deepest red colour), and hence dominate the low-temperature NTE behaviour for both Zn(CN)$_2$ and Cd(CN)$_2$.}
    \label{fig2}
\end{figure*}

At low temperatures, the ostensibly very different material silicon also shows NTE \cite{Gibbons_1958}.  The Si--Si bond network traces the same diamondoid structure of the M(CN)$_2$ family, but there is of course no possibility of mapping onto the polyhedral rotations discussed above. Instead, the NTE behaviour of silicon is understood in terms of angular deformations of the Si--Si bond network \cite{Xu_1991}. The undistorted structure is maximally expanded in the sense that any change to its internal angles necessarily results in a volumetric contraction. Hence the transverse acoustic branches, which capture shear-like distortions of the network, both have NTE character [Fig.~\ref{fig1}(b)] \cite{Shah_1972}. This is essentially a geometric effect, the magnitude of which depends on the relative strengths of bond-bending and bond-stretching terms \cite{Xu_1991}. 

In this Letter, we establish an unexpected mapping between these two contrasting NTE mechanisms. We first use \emph{ab initio} density functional theory (DFT) calculations to (re)determine the phonon relations of M(CN)$_2$ networks, confirming the importance of low-energy translational modes in effecting NTE. Neutron total scattering measurements of Cd(CN)$_2$ then allow us to derive experiment-driven models of local atomic displacements in this material and so to test the validity of the `skipping-rope' mechanism illustrated in Fig.~\ref{fig1}(a). Anticipating our results, we find instead that M atom displacements are essentially as large as those of the C and N atoms. Hence there is a natural mapping of each M--CN--M linkage onto a single M$\ldots$M connection with the cyanide ion acting as a pseudo-spring. Projecting the M(CN)$_2$ phonon dispersion onto M-atom displacements reveals a hidden phonon spectrum closely related to that of Si. The translational NTE branches of M(CN)$_2$ have the same character as the NTE shear modes in Si and we demonstrate that the simple geometric rules used to rationalise the extent of NTE in the latter apply also to M(CN)$_2$. Our analysis highlights the conceptual power of abstracting the lattice dynamics of network structures in terms of their underlying `framework modes'---an approach of potential importance across a wide range of different materials.

\emph{Identifying the NTE phonon modes}---Our starting point was to use \emph{ab initio} density functional theory (DFT) calculations to determine the phonon dispersion relations for Zn(CN)$_2$ and Cd(CN)$_2$. The extent to which individual modes influence thermal expansion is quantified by the corresponding mode Gr\"uneisen parameters \cite{Gruneisen_1926,Misquitta_2013, Wang_2022}
\begin{equation}
    \gamma(\nu, \mathbf k) = -\frac{{\rm d}\ln\omega(\nu, \mathbf k)}{{\rm d}\ln V},
\end{equation}
which capture the sensitivity of the angular frequency ($\omega$) of a mode ($\nu$) at each wave-vector ($\mathbf k$) to changes in crystal volume ($V$). Modes with $\gamma < 0$ soften at lower volumes and favour NTE, whilst modes with $\gamma > 0$ drive PTE behaviour. Our results [Fig.~\ref{fig2}] are consistent with previous calculations \cite{Misquitta_2013, Zwanziger_2007, Wang_2022} and identify three main NTE branches. The lowest-energy branch at $\omega<5$\,meV, for which $\gamma$ is most negative, is doubly degenerate and has translational character; the higher-energy branches at $\omega\sim15$\,meV and $30<\omega<40$\,meV are each triply degenerate and correspond to correlated M(C/N)$_4$ rotations \cite{Misquitta_2013,Goodwin_2006}. While the general form of the phonon dispersion is similar for Zn(CN)$_2$ and Cd(CN)$_2$, the energy scale is reduced in the latter due to the larger mass of Cd relative to Zn, as noted elsewhere \cite{Zwanziger_2007}.

The low energies and large negative Grüneisen parameters of the translational branches mean that these modes dominate the NTE behaviour, especially at low temperatures.

\begin{figure}
    \includegraphics{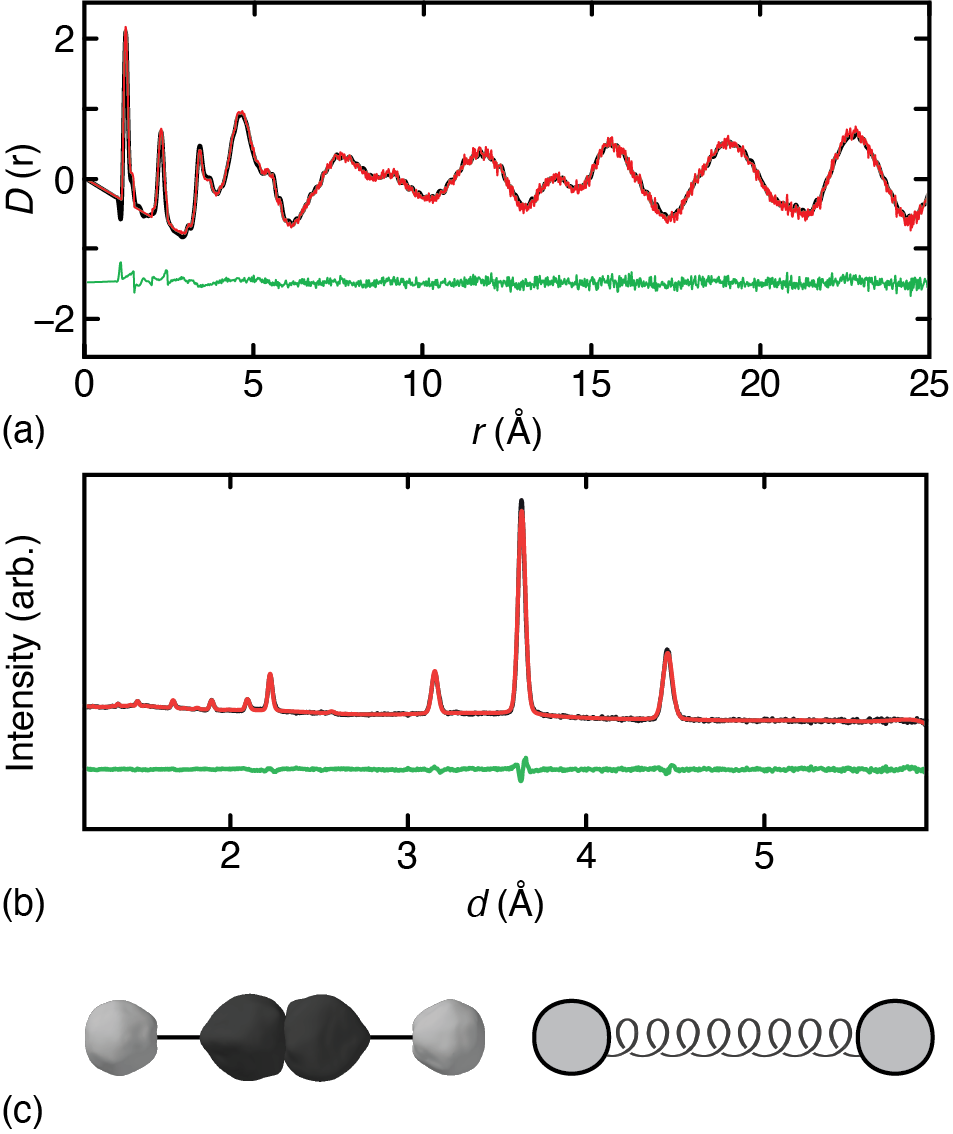}
    \caption{Experimental neutron total scattering data for Cd(CN)$_2$ at 300\,K and corresponding RMC fit. (a) The neutron pair distribution function, with data shown in black, RMC fit in red, and difference function (fit $-$ data) shown in green. (b) Reciprocal-space scattering function, shown here as a conventional Bragg diffraction pattern, with line colours as in (a). (c) Projection of the atomic positions within Cd--C/N--C/N--Cd linkages onto a single common coordinate system. The symmetrised distributions of projected atoms (left) are shown as 90\% isosurfaces and reflect the degree of local thermal displacements of Cd (grey), and C and N atoms (black), deconvolved of contributions from inter-framework slipping \cite{Coates_2021}. That all atom types show similar thermal distributions suggests the CN linkages behave as pseudo-springs connecting neighbouring Cd atoms (right).}
    \label{fig3}
\end{figure}

\emph{Local distortions from experiment}--- To test the validity of the NTE mechanism illustrated in Fig.~\ref{fig1}(a), we used a combination of neutron total scattering measurements and reverse Monte Carlo (RMC) refinements to develop a real-space interpretation of vibrational motion in M(CN)$_2$ materials. There are some peculiarities of the M(CN)$_2$ structure type that informed this particular choice. Both Zn(CN)$_2$ and Cd(CN)$_2$ consist of a pair of interpenetrated diamondoid networks, with a corresponding crystal symmetry ($Pm\bar3n$) that enforces identical distances between neighbouring M$\ldots$M or C/N$\ldots$C/N pairs both within and between networks. Hence direct interpretation of local structure measures, such as the pair distribution function (PDF), is complicated by the superposition of contributions arising from thermal motion within a single network and from the displacements of one network relative to the other. RMC is well placed to disentangle these two contributions because it fits the PDF using an atomistic representation that is interpretable as a structural snapshot of the system undergoing vibrational motion.

Our experimental neutron scattering data for isotopically-enriched $^{114}$Cd(CN)$_2$ \cite{Coates_2018,Coates_2021}, measured at 300\,K using the POLARIS instrument at ISIS \cite{POLARIS}, are shown in both real- and reciprocal-space representations in Fig.~\ref{fig3}(a,b). The Bragg intensities place a constraint on the average extent of vibrational motion through the Debye--Waller factors (single-particle correlation functions), and the PDF adds information regarding correlated displacements via its sensitivity to two-body correlations \cite{Jeong_1999,Jeong_2003,Goodwin_2005b}. RMC refinements against these data made use of the RMCProfile code \cite{Tucker_2007}, and were carried out using $8\times8\times8$ supercells of the $Pn\bar3m$ Cd(CN)$_2$ structure, with cyanide ion orientations determined according to the structural spin-ice model presented in Ref.~\citenum{Coates_2021}. Further details of the refinement approach and parameters used are given as supporting information. The final RMC model gave an excellent fit to data [Fig.~\ref{fig3}(a,b)].

With access to an atomistic representation of the vibrational motion in Cd(CN)$_2$, we were able to extract from our RMC configurations the degree of local vibrational motion across individual Cd--C--N--Cd linkages. For each such linkage, we identified the local Cd$\ldots$Cd axis expected from the high-symmetry average structure. A suitable geometric transformation was then applied to bring each such axis into coincidence, allowing us to project the displacements of all Cd--C--N--Cd linkages onto a single collective distribution. The distribution so obtained is represented in Fig.~\ref{fig3}(c). Were local vibrational motion to be dominated by the skipping-rope mechanisms of Refs.~\citenum{Kepert_2005,Chapman_2005,Misquitta_2013}, we would expect to see a greater distinction between the transverse displacements of the C/N atoms and the isotropic displacements of the anchoring Cd atoms than is evident in our RMC refinements. Instead the magnitude of Cd, C, and N displacements in this local reference frame is roughly isotropic and surprisingly similar from one atom type to another. Hence the picture developed in Fig.~\ref{fig1}(a) (and elsewhere)---as appealing as it is---does not accurately capture the key microscopic behaviour of M(CN)$_2$. 

\begin{figure*}
    \centering
    \includegraphics{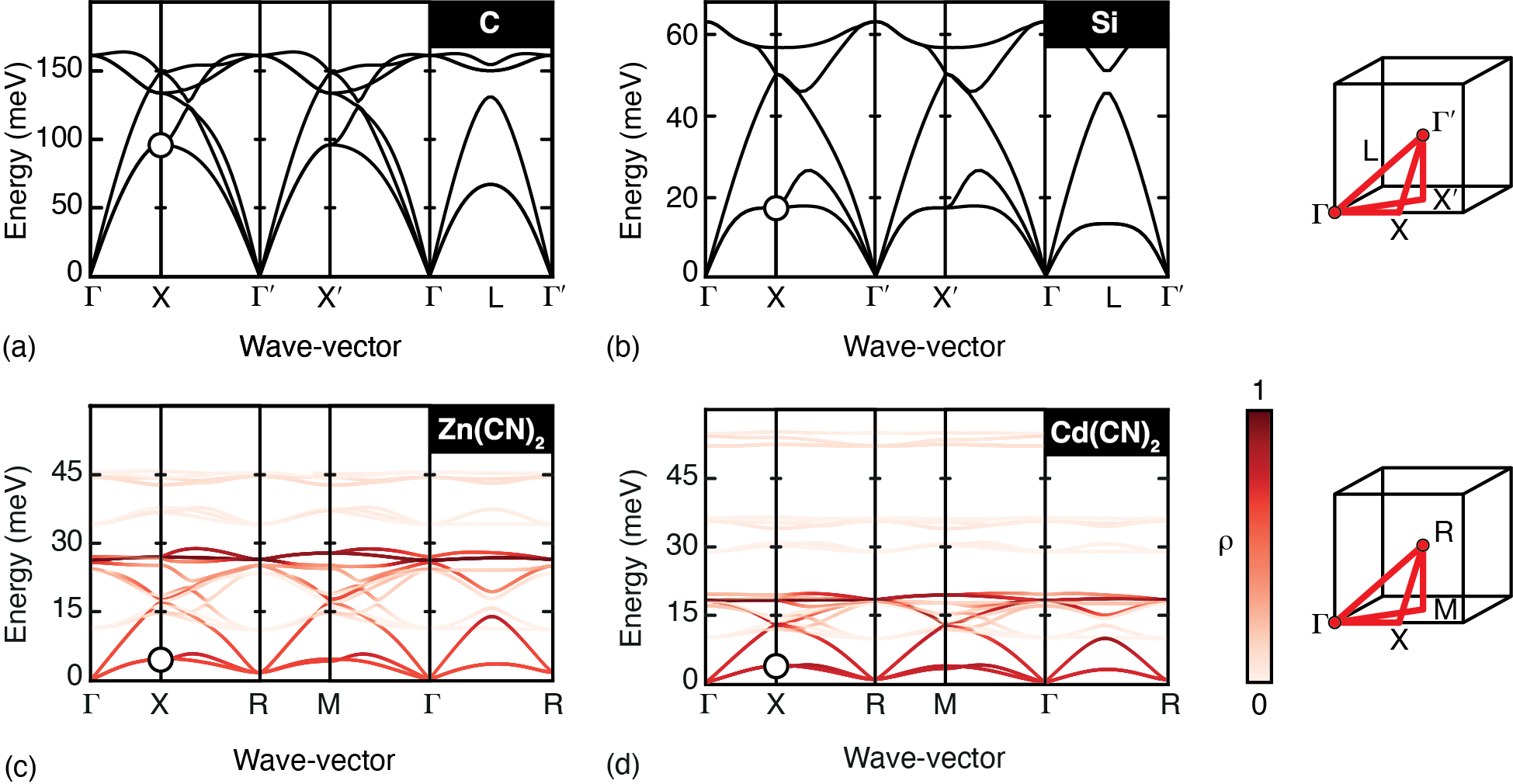}
    \caption{Phonon dispersions of diamond and silicon compared to the framework-weighted phonon dispersion curves of Zn(CN)$_2$ and Cd(CN)$_2$. The diamond (a) and silicon (b) phonon dispersion curves are plotted along a similar path of reciprocal space to that of the M(CN)$_2$ systems. For Zn(CN)$_2$(c) and Cd(CN)$_2$ (d), each mode is coloured according to the relative displacements of the M atoms, capturing the framework modes of M(CN)$_2$. Note the similar form amongst the four sets of dispersion curves. The zone-boundary TA modes discussed in the text are highlighted using open circles. These modes involve angular deformations of the diamondoid lattice as shown in Fig.~\ref{fig1}(b).}
    \label{fig4}
\end{figure*}

\emph{Framework modes}---The distribution of Fig.~\ref{fig3}(c) suggests that the dominant vibrational motion might be better understood in terms of collective displacements of the entire M--CN--M linkages. The interpretation is not that this linkage is rigid (\emph{e.g.}\ as in the `rigid-rod' model for NTE in Cu$_2$O \cite{Sanson_2006}), but rather that the degrees of freedom of the C and N atoms can be integrated out such that the cyanide ion behaves as if it were a pseudo-spring acting between connected M-atom nodes. To test this interpretation, we projected the calculated M(CN)$_2$ phonon dispersion curves onto the contribution from the M-atom displacements alone, using a normalised weighting $0\leq\rho(\nu,\mathbf k)\leq1$ that quantifies the magnitude of M-atom displacements for each phonon mode with label $\nu,\mathbf k$ (see SI).  The displacement-weighted phonon dispersion curves for Zn(CN)$_2$ and Cd(CN)$_2$ are shown in Fig.~\ref{fig4}; we refer to the coarse-grained vibrational modes these characterise as the `framework modes' of the system. Importantly, the key low-energy NTE branch of both Zn(CN)$_2$ and Cd(CN)$_2$ is preserved in this mapping---hence the dominant NTE mechanism ought to be understandable in terms of the behaviour of the diamondoid framework itself, independent of the cyanide-ion displacements (\emph{i.e.}\ neglecting correlated polyhedral tilts or skipping-rope motion altogether).

\emph{The diamond structure is key}---The framework mode dispersion relations of M(CN)$_2$ are closely related to those of diamond and silicon [Fig.~\ref{fig4}], albeit that the presence of two interpenetrating frameworks in M(CN)$_2$ changes the symmetry of reciprocal space between the two families. The overall energy scale of the dispersion is understandably much lower in the case of the cyanide frameworks: it is quite intuitive that M$\ldots$M linkages would be much softer than C--C or Si--Si bonds. But the form of the dispersion also varies in the \emph{relative} energy of the transverse acoustic (TA) branches with respect to the optic branches. It is the TA branches that correspond to the NTE shear modes of the diamond structure [Fig.~\ref{fig1}(b)], and their zone-boundary energies provide a measure of the angular rigidity of the network.

The formalism of Ref.~\citenum{Xu_1991} for diamondoid systems gives a geometric estimate of the Grüneisen parameter at the Brillouin zone boundary (X-point):
\begin{equation}
\gamma_{\rm geom}=\frac{2}{3} - \frac{8}{\sqrt{3}}\frac{B_{0}d_{0}}{m\omega^2_{0}}.\label{geom}
\end{equation}
Here $B_0$ is the bulk modulus, $d_0$ the inter-node separation, $m$ the node mass, and $\omega_0$ the zone-boundary TA angular frequency (X point). The last term of Eq.~\eqref{geom} can be interpreted as the ratio of restoring forces for compressive (numerator) and angular (denominator) deformations. The value of $\gamma$ is most large and negative when the latter is very much smaller than the former. We list in Table~\ref{table1} the various terms in Eq.~\eqref{geom} for each of C, Si, Zn(CN)$_2$, and Cd(CN)$_2$, together with the calculated values of $\gamma_{\rm geom}$. We find excellent agreement with the DFT Gr{\"u}neisen parameters. Hence the contribution of the transverse acoustic branches to NTE can be understood essentially entirely in terms of the geometric effect illustrated in Fig.~\ref{fig1}(b). In particular, there is no need to invoke mechanisms of polyhedral rotations or transverse vibrational motion of the C and N atoms other than through their role in allowing the M--C--N--M linkage to behave as if a pseudo-spring.

\begin{table}
\centering
\caption{Parameters used in the geometric estimation of NTE mode Grüneisen parameters [Eq.~\eqref{geom}] and comparison with DFT results. \label{table1}}
\begin{tabular}{l|cccc}
& C & Si & Zn(CN)$_2$ & Cd(CN)$_2$ \\ \hline
$B_0$ / GPa & 442 & 98.7 & 36.9 & 33.2 \\
$d_0$ / \AA & 1.54 & 2.35 & 5.12 & 5.46 \\
$m$ / g\,mol$^{-1}$ & 12.0 & 28.1 & 117 & 164 \\
$\omega_0$ / THz & 146 & 25.6 & 6.79 & 5.47 \\\hline
$\gamma_{\rm geom}$(X) & $-$0.08 & $-$2.8 & $-$9.0 & $-$9.6 \\
$\gamma_{\rm DFT}$(X) & 0.20 & $-$1.9 & $-$9.9 & $-$9.2 \\
\hline
\end{tabular}
\end{table}

As observed in previous studies, our DFT values for the bulk moduli and zone-boundary TA frequencies of M(CN)$_2$ frameworks are overestimated with respect to experiment \cite{Zwanziger_2007,Ding_2008}; this discrepancy is greater for the case of M = Cd, where thermal reorientation of cyanide ions play a role in the lattice dynamics \cite{Coates_2021}, than it is for Zn. Hence we do not attach too great an importance to the numerical values of $\gamma$, except to note that they are internally consistent and are much larger in magnitude for M(CN)$_2$ than for C or Si as a result of the relative stretching and angular deformation energies noted above. In fact, working within the formalism of Eq.~\eqref{geom}, the experimental values of $B_0$ (33.4\,GPa \cite{Fang_2013} and 13.6\,GPa \cite{Coates_2023}) and $\omega_0$ (3.7\,THz \cite{Chapman_2006} and 1.9\,THz) for Zn(CN)$_2$ and Cd(CN)$_2$, respectively, predict an even more extreme $\gamma_{\rm geom}\simeq-30$. These large and negative Grüneisen parameters for zone-boundary TA modes are consistent with the observation of dynamic instabilities under relatively modest pressures; while the high-pressure structure of Cd(CN)$_2$ is not known \cite{Coates_2023}, that of Zn(CN)$_2$ involves framework shear of the kind shown schematically in Fig.~\ref{fig1}(b) \cite{Collings_2013}.

\emph{Concluding remarks}---The key result of our study is to clarify the most appropriate conceptual framework within which to understand the dominant lattice dynamics of M(CN)$_2$ NTE materials. In conventional framework materials---\emph{e.g.}\ silcates, zeolites, and perovskites---the rigid-unit mode (RUM) model usually plays this role \cite{Giddy_1993,Hammonds_1996,Dove_2019,Dove_2023}. But, while cyanides and other molecular frameworks certainly support RUMs \cite{Goodwin_2006}---and indeed the TA branches can be couched in terms of translational RUMs \cite{Misquitta_2013}---the more relevant coarse-graining appears to be simply in terms of the underlying framework connectivity. We use the term `framework modes' to denote the coarse-grained lattice dynamics of the framework itself---\emph{i.e.} nodes connected via pseudo-springs to give an appropriate degree of compressive and angular rigidity. In the diamondoid M(CN)$_2$ structures, these modes include the low-energy branches that dominate NTE. But equivalent modes may be relevant to the low-energy dynamics of other extended network structures, such as MOFs, for which the conventional RUM picture appears to break down \cite{Rimmer_2014}. Common to both approaches is the motivation to simplify the lattice dynamics of otherwise complex materials and to identify the key structural features responsible for phenomena of physical interest. In this spirit, the framework mode approach also serves to identify the possible design principles one might use to optimise NTE in diamondoid networks: the key microscopic feature being to combine compressive rigidity with angular flexibility \cite{Bhogra_2023}.
 
\begin{acknowledgments}
The authors gratefully acknowledge financial support from the E.R.C. (Grant 788144), funding from the EPSRC Centre for Doctoral Training in
Inorganic Chemistry for Future Manufacturing (OxICFM) (EP/S023828/1) (studentship to Q.G.), and the Leverhulme Trust (Grants RPG-2015-292 and RPG- 2018-268). The authors acknowledge ISIS for the neutron beamtime provision. Neutron beamtime at the ISIS Neutron and Muon Source (RB 1720378) was provided by the UK Science and Technology Facilities Council (STFC).  Data files may be obtained from http://doi.org/10.5286/ISIS.E.RB1720378
\end{acknowledgments}

\end{document}